\RequirePackage{fix-cm}
\documentclass[smallextended]{svjour3}       
\smartqed  
\usepackage{graphicx}
\usepackage[protrusion=true,expansion=true]{microtype}				
\usepackage{amsmath}					
\usepackage{amsfonts}
\usepackage{mathabx}									
\usepackage[svgnames]{xcolor}									
\usepackage[hang, small,labelfont=bf,up,textfont=it,up]{caption}	
\usepackage{epstopdf}												
\usepackage{subfig}													
\usepackage{booktabs}												
\usepackage{fix-cm}													
\usepackage{todonotes}
\usepackage{appendix}

\begin{document}

\title{Onsager's irreversible thermodynamics of  the dynamics of transient pores in spherical lipid vesicles}
\author{L. Mart\'inez-Balbuena \and E. Hern\'andez-Zapata \and I. Santamar\'{\i}a-Holek
}

\institute{L. Mart\'inez-Balbuena \at
              FCFM, Universidad Aut\'onoma de Puebla, Pue. 64520, M\'exico. \\
              \email{mtz.luciano@gmail.com}          
           \and
           E. Hern\'andez-Zapata \at
              Dpto. de Recursos de la Tierra, Universidad Aut\'onoma Metropolitana Unidad Lerma, Estado de M\'exico 52006, M\'exico.
            \and 
            I. Santamar\'{\i}a-Holek \at
            UMDIJ-Facultad de Ciencias,  Universidad Nacional Aut\'onoma de M\'exico Campus Juriquilla, Quer\'etaro 76230, M\'exico.
}

\date{Received: date / Accepted: date}

\maketitle

\begin{abstract}
Onsager's irreversible thermodynamics is used to perform a systematic deduction of the kinetic equations governing the opening and collapse of transient pores in spherical vesicles. We show that the edge tension has to be determined from the initial stage of the pore relaxation and that in the final state the vesicle membrane is not completely relaxed, since the surface tension and the pressure difference are about $25\%$ of its initial value. We also show that the pore life-time is controlled by the solution viscosity and its opening is driven by the solution leak-out and the surface tension drop. The final collapse is due to a non-linear interplay between the edge and the surface tensions together with the pressure difference. Also, we discuss the connection with previous models.
\end{abstract}

\section{Introduction}\label{intro}
Biological pores are important entities in cells and play a crucial role in the spatial and temporal control of energy and matter fluxes. For instance, the modulation of membrane permeabilization mechanisms by the physicochemical properties of lipid bilayers is quite important for understanding antimicrobial mechanisms \cite{Matsuzaki,Zasloff}. When membrane organization is intact after transient pore formation, the permeabilization can be partially compensated by ion channels and pumps. When 
membrane organization becomes affected, this may eventually lead to the disruption of the membrane. In this case, the survival of the cell depends crucially of the speed with which the membrane is repaired \cite{McNeil(2003)}.  In general terms, pores in biological membranes have different origins and their mechanisms of formation may depend upon proteins and peptides \cite{Zasloff,Papo}. In the literature, it is suggested that transient pores are more commonly formed after the interaction between peptides and the lipid membranes \cite{Matsuzaki,Zasloff}. Nevertheless, pore formation has also been studied in prebiotic environments, that is, without peptides and proteins, with the aim to understand their role as a transport mechanism in protocells modeled by unilamellar vesicles \cite{Sakuma}. Exocytosis and endocytosis involve processes of membrane fusion and fission giving rise to the  formation of a \emph{fusion pore}, a channel, through which secretions are released from the vesicle to the cell exterior  \cite{Vardjan,Suchmita(1997),GaryJ(2009),APicco(2015),MonckJR(1994),PaladeG(1975)}. The  formation of the \emph{fusion pore} occurs through the action of protein machinery (SNAREs) that have to apply mechanical forces in the range of 2-20pN for the opening of the pore \cite{Bykhovskaia}. Pores are also important for many technological applications such as nanomedice, sensing, and nanoelectronics \cite{Majd(2010)}.

Biological membranes are mainly composed of lipid molecules that form bilayers due to their amphiphilic structure. Lipid membrane are easily used in experiments to form uni- and multilamellar vesicles of different shapes and therefore they are very good models to investigate the main features of the formation and collapse of pores. In the simplest case, the pore formation and growth are generated by physical forces and do not require the presence of proteins. This process can be modeled by liposome rupture due to swelling. 
 
Recently, giant unilamellar vesicles immersed in an aqueous solution and composed of a single phospholipid bilayer and typical radius, $R \sim 10 - 100\mu$m, have been used to study the formation and collapse of transient pores  \cite{OSandre(1999),ErdemK(2003),ThomasP(2010),KarinA(2005)}.  The rupture and a transient pore appear as response to mechanics stress over the membrane. Once the mechanical stresses along the membrane exceed the lysis tension, the vesicle ruptures and a transient pore may appear \cite{ThomasP(2010)}. 
Application of intense optical illumination on vesicles containing fluorescent membrane probes induces a sudden increase of the membrane surface tension until the creation of a pore in the bilayer \cite{OSandre(1999),ErdemK(2003),NRodriguez(2006)}. Another technique used for creating pores in vesicles is the adhesion of the vesicle on an attractive surface \cite{OSandre(1999)}. 
More recently, Dimova and coworkers \cite{ThomasP(2010),KarinA(2005)} have developed a different method of inducing pores in giant unilamellar vesicles by the application of electric pulses. The dynamics of the pore is then captured by means of a fast imaging digital camera with a high temporal resolution. This method avoids some disadvantages of previous techniques, such as the use of fluorescent probes and glycerol to increase the solution viscosity, and allows better resolution of pore dynamics. For instance, the initial pore size is not well documented in all cases except in Ref. \cite{ThomasP(2010)}, for which the initial pore size is about $r_{min} = 0.75 \mu m$.
However, for vesicles with similar diameters,  all these techniques produce pores with maximum radii in the range $r_{max} = 5 - 11 \mu m$.

In this work we are interested in the theoretical description of pore dynamics after its formation. This dynamics can be divided into three stages. The first stage is characterized by rapid growth of the pore caused by the excess surface tension and the difference between the inner and the external pressures. The second stage is a relaxation process in which the slow shrinking of the pore is dominated by the edge tension. Finally, at long times the rapid pore collapse is due to a non-linear interplay among the main forces participating in the dynamics, the edge and surface tensions together with the pressure difference.

The evolution of pores in membranes has been previously described in Ref. \cite{Brochard(2000)} by adopting a model coming from the theory of viscous bare films in which pore opening is governed by the transfer of surface energy to viscous losses, and edge forces are neglected \cite{Debregeas(1995),Diederich(1998)}. An important assumption of the mentioned description is that the solution viscosity is small and the membrane viscosity controls the closing process. However, this seems to be in contradiction with the experimental fact that, in the case of membranes, the solution viscosity can be used to regulate the pore closing velocity \cite{ErdemK(2003),ThomasP(2010)}. An extension of this theory was formulated in Ref. \cite{RRyham(2011)} where a lateral stress on the pore arising from the tangential movement of viscous aqueous solution relative to the membrane was included. This correction introduces a dependence on solution viscosity that accounts for the dependence of the pore closing velocity on the solution viscosity.  As a consequence, the dependence of the pore collapse kinetics on the membrane viscosity becomes negligible in the initial stages. However, the final stage depends crucially on membrane viscosity. Other approaches taking into account the shape of the vesicle as a variable were proposed in \cite{RRyham(2012)}.
 
In this article we use Onsager's irreversible thermodynamics to deduce a model accounting for the kinetics of pore collapse in membranes. We begin by evaluating the free energy differential of an open vesicle, including the bending energy, the surface tension, the edge tension, and the coupling of the vesicle volume to its environment through a pressure difference term. Using the second law of thermodynamics we are able to deduce the kinetic equations that govern the time evolution of the pore and vesicle radii. By imposing consistent physical conditions on the resulting equations, we propose the correct dependence of the Onsager phenomenological coefficients on the pore and the vesicle radii. The obtained equations are numerically solved showing excellent comparison with experiments and a consistent behavior. In this way, we show that the closing velocity is mainly controlled by the edge tension and the solution viscosity. We also determine the time evolution of the surface tension and of the pressure difference. Our results question previous interpretations of the collapse dynamics. Previous models can be deduced from our general formalism depending on the election of the Onsager phenomenological coefficients. The corresponding inconsistencies are discussed. 

The article is organized as follows. Section \ref{sec:1} is devoted to discussing the expression for the free energy differential of an open vesicle using thermodynamics arguments. Then we use the free energy differential and Onsager's thermodynamics to derive the dynamical equations describing the process in Section \ref{sec:2}. Finally, in Section \ref{sec:concl}, we discuss our results and present our conclusions.

\section{Thermodynamics of a stressed vesicle}\label{sec:1}
\begin{figure}[t!]
\centering
\includegraphics[width=4.40in]{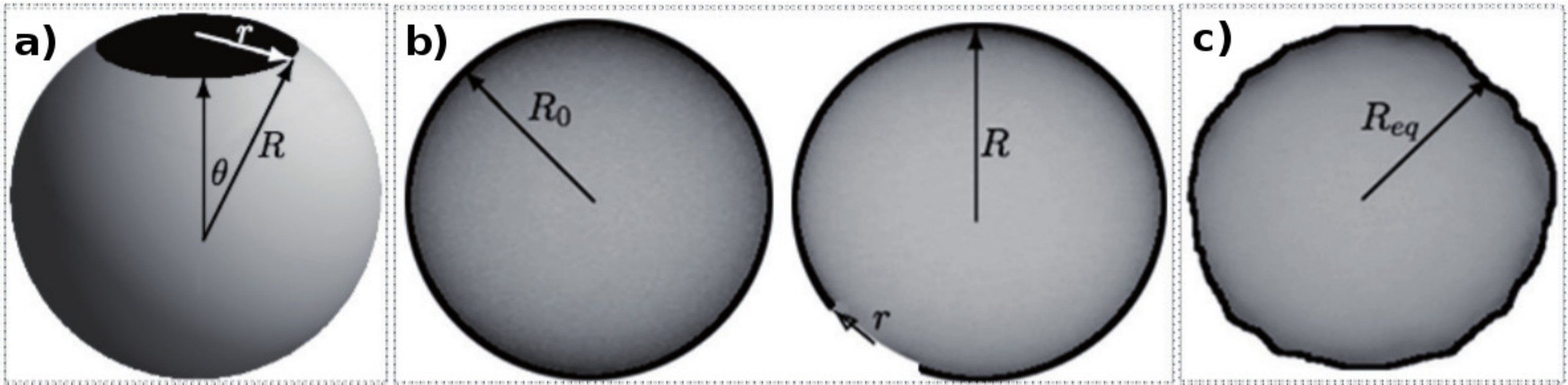}
 \caption{{\small a) Schematic representation of a vesicle with pore, b) Stressed vesicle and formation of a transient pore. c) Relaxed vesicle.}}
\label{fig:fig01}
\end{figure}

Consider a relaxed vesicle radius $R_{eq}$ immersed in an aqueous solution at constant temperature $T$ and total volume $V=V_ {in}+V_ {out}$ with $V_ {in}=(4/3)\pi R_{eq}^{3}$ the volume inside the vesicle and $V_ {out}$ the volume outside the   vesicle, see Fig. \ref{fig:fig01} c). The curved membrane has by definition a bending free energy per unit area given by $\kappa_{b}/R_{eq}^{2}$, with $\kappa_{b}=2\kappa+\bar{\kappa}$ being $\kappa$ and $\bar{\kappa}$ the bending and the saddle-splay moduli, respectively \cite{WHelfrich(1973),EHernandez(2009)}. Furthermore, due to the presence of the curved membrane, a finite pressure difference between the inside and outside of the vesicle exists, $\Delta P = P_{in}-P_{out}>0$.

After an initial perturbation that increases the membrane surface tension to $\sigma_0$ and the vesicle radius to $R_0$, a pore radius $r(t)$ is produced that initiates a time dependent process in which the radius of the vesicle becomes a function of time, $R(t)$, see Fig. \ref{fig:fig01}b. Here, we will analyze the dynamical process of the pore growth and collapse in terms of $r(t)$ and $R(t)$.

The free energy change of the system has four contributions: $dF=dF_{V}+dF_{\sigma}+dF_{B}+dF_{l}$, coming from volume ($F_{V}$), surface tension ($F_{\sigma}$), bending curvature ($F_{B}$) and edge tension ($F_{l}$). Thus, in the stressed state we have that the free energy has the following expression \cite{EHernandez(2009)}
\begin{equation}
dF=-\Delta PdV_{in}+\tilde{\sigma}
dA+\gamma dl,
\label{eq:ec-dF}
\end{equation}
where at the right hand side we used the relation $dV_ {out} =-dV_ {in}$ and $\tilde{\sigma}$ stands for the total effective surface tension given by $\tilde{\sigma} = \sigma+ \kappa_{b}/R^{2}$. The last term, responsible for pore closure is the contour free energy with $\gamma$ the edge tension and $l=2\pi r$ the pore contour length. In this form, for small pore radius ($r(t)/R(t)\ll1$) the open vesicle volume and area may be written as $V_{in} = 4\pi R^3/3$ and $A=4\pi R^{2}-\pi r^{2}$ respectively, thus allowing us to express the free energy differential in terms of radii $r$ and $R$:
\begin{equation}
{dF}= 4\pi R\left(2\tilde{\sigma}-R\Delta P\right){dR}+2\pi \left(\gamma -\tilde{\sigma}r\right){dr}.
  \label{eq:ec_dotF-bis}
\end{equation}
Notice that before the initial perturbation that yields the appearance of a surface tension ($\sigma =0$) in the membrane of the vesicle, the radius of the vesicle is $R_{eq}$. In that case, Eq. (\ref{eq:ec_dotF-bis}) reduces to:
$\Delta P_{eq} = {2\kappa_{B}}/{R_{eq}^{3}}$.

The explicit expression of the surface tension can be deduced by noticing that the energy required to stretch a membrane by reducing the thermal fluctuations on the vesicle shape is given by the expression \cite{YLevin(2004)}
\begin{equation}
F_{\sigma}=\sigma_{c}\dfrac{\left(A-A_{eq}\right)^{2}}{2A_{eq}},
\label{eq:ec-FS}
\end{equation} 
where $\sigma_{c}$ is a characteristic surface tension, while $A_{eq}$ and $A$ are vesicle areas before and after increasing the membrane tension. Since the temperature and the volume of the solution are controlled, the change in surface energy  $dF_{\sigma}$ with respect to the area of the membrane is \cite{PKondepudi(2007),YLevin(2004)}
 \begin{equation}
\sigma=\left(\frac{\partial F_{\sigma}}{\partial A}\right)_{T,V}= \sigma_{c}\left(\dfrac{A}{A_{eq}}-1\right).
\label{eq:ec-Sig}
\end{equation}
Using the explicit expression of the area in terms of $r$ and $R$ the last expression takes the following form
\begin{equation}
\sigma= \sigma_{c}\left[\dfrac{R^{2}}{R_{eq}^{2}}-\dfrac{r^{2}}{4R_{eq}^{2}}-1\right]. 
\label{eq:ec_SigAp}
\end{equation}
Equation (\ref{eq:ec_SigAp}) allows one to calculate the evolution of the surface tension during the entire process of opening and collapse of the pore. A similar expression was used in  \cite{Brochard(2000)} to express the area occupied by the lipid when the membrane is stretched.

After the pore collapse, the final radius $R_{f}$ of the vesicle may be, in general, slightly larger than $R_{eq}$ (see Ref. \cite{ErdemK(2003)} and Figure 2 below) and therefore the final value of the pressure difference contains an additional contribution, with respect to $\Delta P_{eq} $, coming from a residual surface tension
\begin{equation}
\Delta P_{f} = \dfrac{2\tilde{\sigma}_{f}}{R_{f}}=\dfrac{2\sigma_{f}}{R_{f}}+\dfrac{2\kappa_{B}}{R_{f}^{3}}.
\label{eq:ec_LaplaceP}
\end{equation}
This equation states that the Laplace relation gives the final pressure with a correction term due to the bending energy of the vesicle. For a lipid bilayer, the value of the surface tension just before the opening of the pore takes the value $\sigma_{0}\sim 10^{-5}N/m$ whereas the bending modulus is about $\kappa_{b}\sim10^{-20}J$. This implies that the bending modulus is $\kappa\sim10^{-20}J$ at 18 degrees Celsius \cite{WRawicz(2000)} whereas the saddle-splay modulus is approximately given by $\bar{\kappa}\sim-a\kappa$ with $a\sim 1$ or less \cite{TDLe(2000),DPSiegel(2004)}. If the vesicle size is smaller than a critical size, $R_{eq}<(\kappa_{b}/\sigma_{eq})^{-1/2}\sim10^{-2}\mu$m, then the correction term due to the bending energy of the vesicle cannot be ignored. In the case of giant unilamellar vesicles, the bending energy effect on the final pressure is small. 

The pressure difference $\Delta P$ decreases as the pore collapses and the radius of the vesicle decreases  because a solvent outflow. Previous simulations of this system indicate that during the collapse the vesicle shape  remains approximately spherical \cite{RRyham(2012)}. Hence, an expression for the pressure difference can be obtained by calculating the mass outflow per unit time for a sphere with a pore which can be expressed in the general form
\begin{equation}
Q=\int{\rho\vec{v}\cdot d\vec{s}},
\label{eq:ec_Q}
\end{equation}
where $\rho$ is the solvent mass density, $\vec{v}$ is the flux velocity crossing the pore with differential surface area $d\vec{s}$. Using the Navier-Stokes equation for stationary states and incompressible fluids, $\eta_{s}\nabla^{2}\vec{v}-\nabla P=0$, the approximate expression for the velocity in this case is
\begin{equation}
v\simeq \frac{2r\Delta P}{\eta_{s}}.
\label{eq:ec_v}
\end{equation}
Considering cylindrical coordinates and combining equations (\ref{eq:ec_Q}) and (\ref{eq:ec_v}), we obtain the following formula for the pressure difference in terms of the mass outflow per unit time
\begin{equation}
\Delta P=\frac{3\eta_{s}}{4\pi \rho r^{3} } Q(R). 
\label{eq:ec_QDP}
\end{equation}
An explicit relation in terms of the vesicle radius will be given in the next section.

\section{Onsager's irreversible thermodynamics and the dynamic equations}\label{sec:2}
The equations governing the dynamics for $r(t)$ and $R(t)$ can be deduced following Onsager's irreversible thermodynamics \cite{PKondepudi(2007)} that establishes phenomenological relations for the time evolution of the system thermodynamic variables in terms of the conjugated generalized forces \cite{PKondepudi(2007)}. This procedure is systematic because it uses the second law of the thermodynamics through the calculation {\bf of} the entropy production.

From thermodynamics, it can be shown the general relation: $dF=-T d_iS < 0$, where $d_iS$ is the entropy produced during an irreversible process occurring at constant volume and temperature \cite{PKondepudi(2007)}. Therefore, the entropy produced per unit time is proportional to the time change of the system Helmholtz free energy
\begin{equation}
T \frac{d_iS}{dt} = - \frac{dF}{dt} = - \frac{\partial F}{\partial r} \frac{dr}{dt} -\frac{\partial F}{\partial R}\frac{dR}{dt} >0.
\label{eq:ec-Entropy}
\end{equation}
In this expression, the time derivatives play the role of flows, $J_{k}={d\zeta_{k}}/{dt} $, whereas the partial derivatives play the role of generalized forces $X_{k}=-\partial F/\partial\zeta_{k}$, \cite{PKondepudi(2007)}. Following the rules of non-equilibrium thermodynamics, we may assume that the flows are linear functions of the forces $J_{k}=\sum_{j}{L_{kj}X_{j}}$, where the $L_{kj}$ are the so-called Onsager's phenomenological coefficients. Additionally, the elements  $L_{kj}$ also obey the Onsager reciprocal relations $L_{kj}=L_{jk}$. These reciprocal relations are associated with cross effects \cite{LOnsager(1931)}. Therefore the linear laws imply
\begin{eqnarray}
\frac{dr}{dt}=-L_{rr}\dfrac{\partial F}{\partial r}-L_{r R}\dfrac{\partial F}{\partial R},\nonumber\\
 \frac{dR}{dt}=-L_{RR}\dfrac{\partial F}{\partial R}-L_{R r}\dfrac{\partial F}{\partial  r}.
  \label{eq:ec_rR}
\end{eqnarray}

The Onsager's coefficients $L_{ij}$ are mobilities characterizing the change of the radii in time (flows) to the change of the energy as a function of the radii (forces). Therefore, the units of the Onsager coefficients are the inverse of Jules-second per square meters: $[L_{ij}^{-1}]=J\,s/m^{2}$. Furthermore, the experimental results \cite{OSandre(1999),ErdemK(2003)} suggest that the duration of the pore dynamics depends on the viscosity of the medium in such a way that the dynamics is slower when the solvent viscosity increases. Since the dimensions of the viscosity are $[\eta_{s}]=J\cdot s/m^{3}$, then we have that 
\begin{equation}
 L^{-1}_{rr}\hspace{0.1cm} \propto\hspace{0.1cm} \eta_{s} l_{1}, \hspace{0.45cm} L^{-1}_{r R}=L^{-1}_{R r}\hspace{0.1cm} \propto\hspace{0.1cm} \eta_{s} l_{2}\hspace{0.3cm} \mbox{and}\hspace{0.3cm}L^{-1}_{RR}\hspace{0.1cm} \propto\hspace{0.1cm} \eta_{s} l_{3},
  \label{eq:ec_OnsagerC}
\end{equation}
with $l_{i}$ ($i=1,2,3$) a characteristic length of the system. In general, the Onsager coefficients may be functions of the state variables $r$ or $R$ \cite{PKondepudi(2007)}. The election of this dependence is crucial and a proper choice leads to a physically consistent description of the process. In the following we will show that pre-existing models can be obtained from our general theory by assuming constant coefficients, whereas a fully consistent system of equations need to consider that the coefficients should depend on the state variables.

The change with time of the free energy given by Eq. (\ref{eq:ec_dotF-bis})  is therefore
\begin{equation}
\frac{dF}{dt}= 4\pi R\left(2\tilde{\sigma}-R\Delta P\right)\frac{dR}{dt}+2\pi \left(\gamma-\tilde{\sigma}r\right)\frac{dr}{dt}.
  \label{eq:ec_dotF}
\end{equation}
In this form, following the general scheme already presented we may write down the following set of dynamical equations for the time evolution of the pore and vesicle radii
\begin{eqnarray}
\frac{dr}{dt}=2\pi L_{r r}\left\{\tilde{\sigma}r-\gamma\right\}-4\pi RL_{r R}\left\{2\tilde{\sigma}-R\Delta P\right\},
\label{eq:ec_dotrp}\\
\frac{dR}{dt}=-4\pi RL_{R R}\left\{2\tilde{\sigma}-R\Delta P\right\}+2\pi L_{R r}\left\{\tilde{\sigma}r-\gamma\right\}.
\label{eq:ec_dotRV}
\end{eqnarray}
Equations (\ref{eq:ec_dotrp}) and (\ref{eq:ec_dotRV}) with a non-zero initial condition for the pore radius predict the dynamics of the transient pore in spherical vesicles. Both equations are coupled through the effective surface tension and the pressure difference terms.

\subsection{Dynamic equations and Onsager coefficients}
Dimensional analysis establishes that the Onsager coefficients have to be inversely proportional to a viscosity and a characteristic length. Irreversible thermodynamics states that phenomenological coefficients can in general be functions of the state variables. Therefore, here we will propose an expression for the Onsager coefficients as functions of the system's state variables $r$ and $R$, that yields a physically consistent model of pore dynamics.

First, we have to emphasize that Eqs. (\ref{eq:ec_dotrp}) and (\ref{eq:ec_dotRV}) are not physically consistent if the Onsager coefficients are constants, because in this case when the pore collapses ($r=0$) the time derivatives $dr/dt$ and $dR/dt$ are proportional to $-\gamma$, which has been considered in the literature as a constant parameter \cite{OSandre(1999),ErdemK(2003),ThomasP(2010),Brochard(2000),RRyham(2011)}. Hence, at $r=0$ the dynamics does not vanishes.
\begin{figure}
  \centering
  \includegraphics[width=1.0\textwidth]{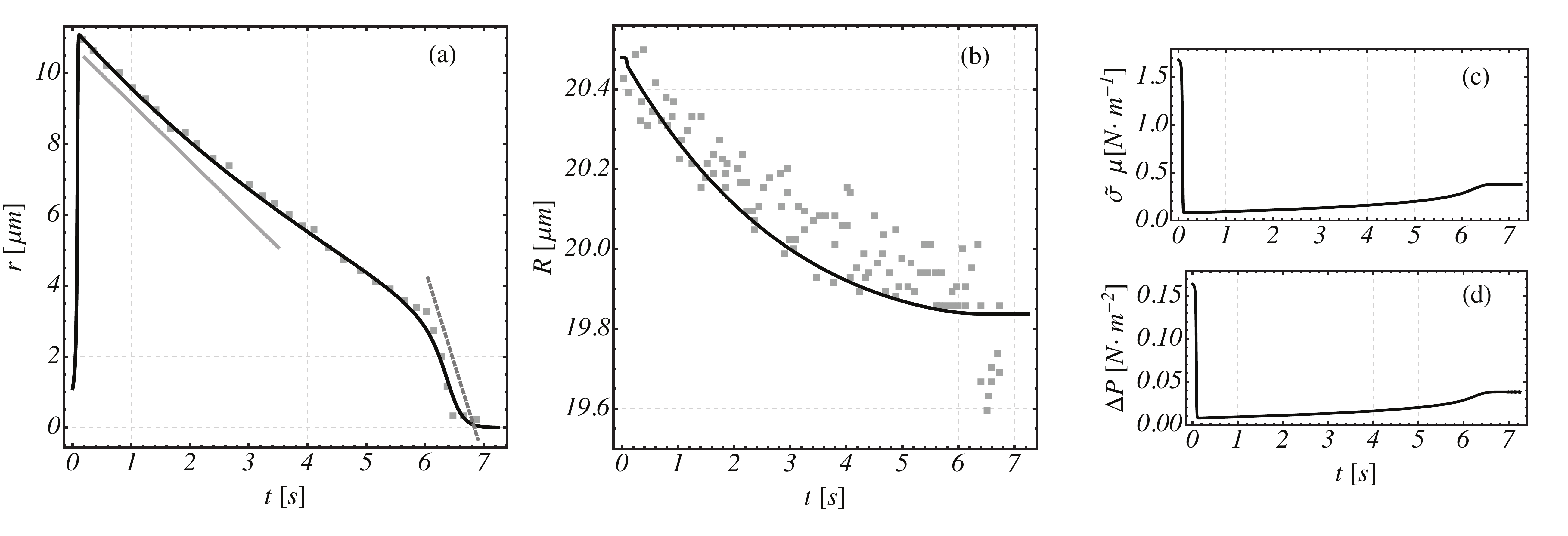}
   \caption{\small Comparison of the numerical solutions (solid lines) of Eqs. (\ref{eq:ec_rpAll}) and (\ref{eq:ec_RVAll}),  along with experimental data (symbols) taken from Ref. \cite{ErdemK(2003)}: (a) pore radius as a function of time, (b) vesicle radius as a function of time, (c) surface tension and (d) pressure difference.  The parameters taken from the experimental measurements are  $R_{eq}=19.6\mu$m, $R_{0}=20.5\mu$m, $r_{0}=1.10\mu$m, $\sigma_{c}=2.9 \times 10^{-5}$N$\cdot$m$^{-1}$, $\eta_{s}=32$cP, and $\kappa_{b}=29 \times 10^{-20}$J. The edge tension was considered as a fitting parameter and was found to be equal to $\gamma=0.92$pN.}
   \label{fig:Dyna_Karatekin}       
\end{figure}

The Onsager coefficients are the inverse relaxation times of the pore dynamics. As implemented in experiments \cite{OSandre(1999),ErdemK(2003),ThomasP(2010)}, changing the solution viscosity 
$\eta_{s}$ allows to slow down the pore dynamics. Thus, we may assume that the Onsager coefficients scale with the solution viscosity $\eta_{s}$. In view of this, a consistent choice of the coefficients that satisfies the conditions (\ref{eq:ec_OnsagerC}), does not introduces unknown free parameters and satisfies conditions $dr/dt=0$ and $dR/dt=0$ at $r=0$, is given by the relations 
\begin{equation}
L_{rr}=L_{RR}=\dfrac{1}{2\pi r_{0}\eta_{s}}\dfrac{r}{R}\hspace{0.7cm}\mbox{ and } \hspace{0.7cm} L_{rR}=\dfrac{1}{2\pi r_{0}\eta_{s}}\dfrac{r^{2}}{R^{2}},
\label{eq:ec_Lij}
\end{equation}
where $2\pi r_{0}$ is the initial length  of the pore (perimeter) in which the initial radius $r_0$ can be estimated from Eq. (\ref{eq:ec_SigAp}). Note that both Onsager coefficients are well defined for all values of the radii. At the initial condition they take the values: $L_{rr}=L_{RR}=1/(2\pi\eta_s R_0)$ and $L_{rR}=r_0/(2\pi \eta_s R^2_0)$, whereas for the final state they vanish. A more general expression for the Onsager coefficients can be proposed by considering the effect of the lipid viscosity 
$\eta_2$ in the form: $L_{ii} \simeq \left({\eta_{s}r_{0} + \eta_2 d}\right)^{-1} \left({r}/{R}\right)$, $L_{ij} \simeq \left({\eta_{s}r_{0} + \eta_2 d}\right)^{-1} \left({r}/{R}\right)^{2}$ with $d$ the thickness of the bilayer. This second term could be of importance in the case of low viscous environments, see Ref. \cite{ThomasP(2010)}.

\begin{figure}
  \centering
  \includegraphics[width=1.0\textwidth]{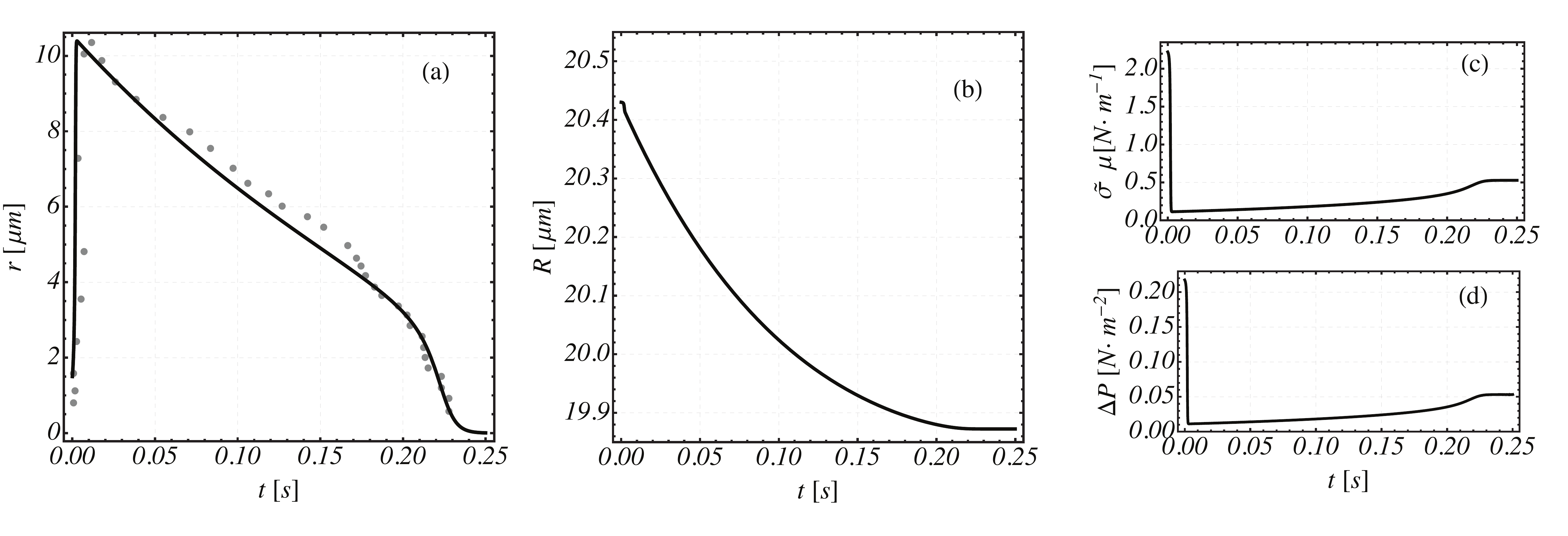}
\caption{\small Comparison of the numerical solutions (solid lines) of Eqs. (\ref{eq:ec_rpAll}) and (\ref{eq:ec_RVAll}),  along with experimental data (symbols) taken from \cite{ThomasP(2010)}: (a) pore radius as a function of time, (b) vesicle radius as a function of time, (c) surface tension and (d) pressure difference.  The parameters taken from the experimental measurements are   $R_{eq}=19.7\mu$m, $R_{0}=20.4\mu$m, $r_{0}=1.15\mu$m, $\sigma_{c}=3.0 \times 10^{-5}$N$\cdot$m$^{-1}$,  $\eta_{s}=1.13$cP,  and $\kappa_{b}=29 \times 10^{-20}$J. The edge tension was considered as a fitting parameter and was found to be equal to $\gamma=1.21$pN.}
\label{fig:Dyna_Dimova}      
\end{figure}

Thus, Eqs. (\ref{eq:ec_dotrp}) and (\ref{eq:ec_dotRV}) take the form
\begin{eqnarray}
r_{0}\eta_{s}\frac{dr}{dt}=\left\{\tilde{\sigma}r-\gamma\right\}\dfrac{r}{R}-2\left\{2\tilde{\sigma}-R\Delta P\right\}\dfrac{r^{2}}{R},
\label{eq:ec_rpAll}\\
r_{0}\eta_{s}\frac{dR}{dt}=-2\left\{2\tilde{\sigma}-R\Delta P\right\}r+\left\{\tilde{\sigma}r-\gamma\right\}\dfrac{r^{2}}{R^{2}}.
\label{eq:ec_RVAll}
\end{eqnarray} 
This set of equations must be complemented with the expressions for the effective surface tension 
$\tilde{\sigma} = \sigma + \kappa_{b}/R^{2}$ with $ \sigma$ given by Eq. (\ref{eq:ec_SigAp}), and the pressure difference $\Delta P$ given by Eq. (\ref{eq:ec_QDP}). The pressure difference $\Delta P$ is related with the time variation of the vesicle volume in the form
\begin{equation}
Q=-\rho\frac{dV_{in}}{dt}=-4\pi\rho R^{2}\dot{R}.
\label{eq:ec_QR}
\end{equation}
From equations (\ref{eq:ec_QDP}) and (\ref{eq:ec_QR}) we obtain
\begin{equation}
\Delta P=-\frac{3\eta_{s} R^{2}}{r^{3}}\dot{R}.
\label{eq:ec_DP}
\end{equation}

\begin{figure}
  \centering
  \includegraphics[width=1.0\textwidth]{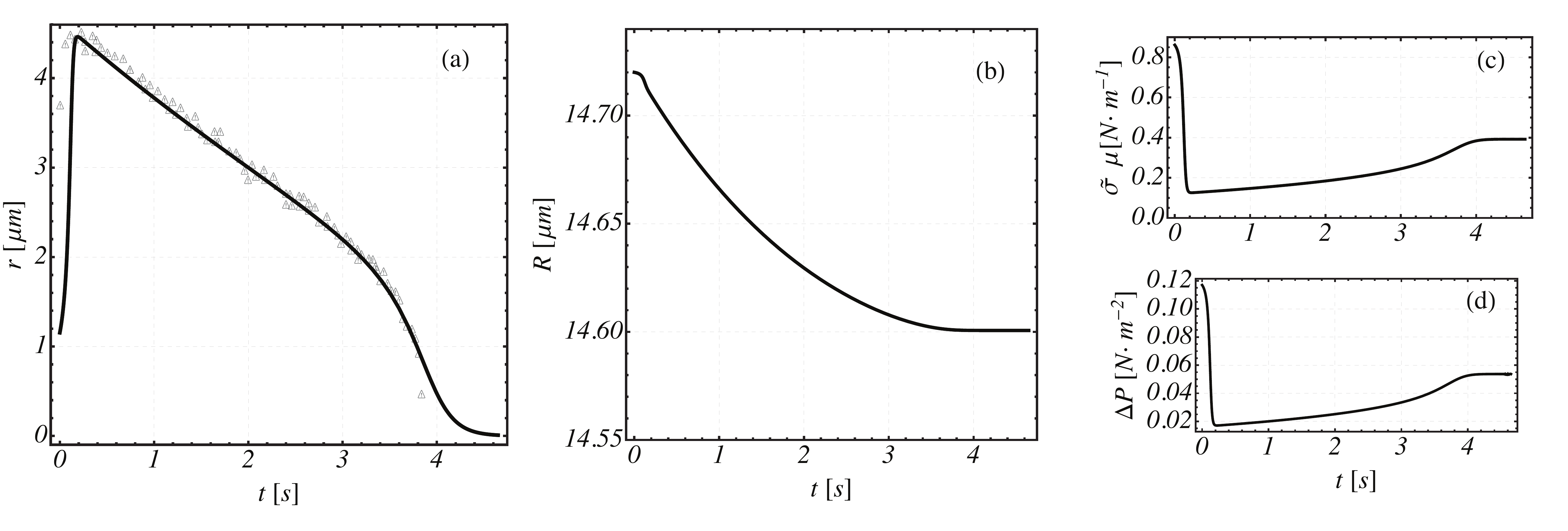}
   \caption{\small Comparison of the numerical solutions (solid lines) of Eqs. (\ref{eq:ec_rpAll}) and (\ref{eq:ec_RVAll}),  along with experimental data (symbols) taken from Ref. \cite{Brochard(2000)}: (a) pore radius as a function of time, (b) vesicle radius as a function of time, (c) surface tension and (d) pressure difference. The parameters taken from the experimental measurements are   $R_{eq}=14.5\mu$m, $R_{0}=14.7\mu$m, $r_{0}=1.15\mu$m, $\sigma_{c}=3.1 \times 10^{-5}$N$\cdot$m$^{-1}$,  $\eta_{s}=32$cP,  and $\kappa_{b}=29 \times 10^{-20}$J. The edge tension was considered as a fitting parameter and was found to be equal to $\gamma=0.57$pN.}
   \label{fig:Dyna_DeGennes}       
\end{figure}

Equations (\ref{eq:ec_rpAll}) and (\ref{eq:ec_RVAll}) constitute a coupled set of two ordinary non-linear differential equations that have to be solved numerically along with (\ref{eq:ec_SigAp}) and (\ref{eq:ec_DP}). 
The results of the numerical integration of the radii $r(t)$ and $R(t)$ (solid lines) are shown in Figures \ref{fig:Dyna_Karatekin}-\ref{fig:Dyna_DeGennes} together with experimental data (symbols) taken from \cite{ErdemK(2003),ThomasP(2010),Brochard(2000)}. Our model reproduces very well the three stages of the dynamics of the pore radius (opening, decay and collapse). In addition, our model allows determining the time dependence of the effective surface tension and of the pressure difference as it is also shown in Figures \ref{fig:Dyna_Karatekin}-\ref{fig:Dyna_DeGennes}. 

The values of the solvent viscosity, $\eta_{s}$, the equilibrium vesicle radius, $R_{eq}$, the initial pore radius, $r_{0}$ and the initial vesicle radius, $R_{0}$ indicated in Table \ref{tab:Table1} and used in Figures \ref{fig:Dyna_Karatekin}, \ref{fig:Dyna_Dimova} and \ref{fig:Dyna_DeGennes},  were taken from experimental reports from Refs. \cite{ErdemK(2003)}, \cite{ThomasP(2010)}, and \cite{Brochard(2000)}, respectively. According to Ref. \cite{WRawicz(2000)}, we assumed that the bending modulus is equal to $\kappa_{b}=29\times10^{-20}$J in all cases. The characteristic surface tension was estimated by using the expression $\sigma_c ={(48\pi\kappa^{2})}/{(R_{eq}^{2}K_{B}T)}$, \cite{Brochard(2000),RRyham(2011)}. The last column in Table \ref{tab:Table1} reports the values of the edge tension obtained from the fits.

In order to conclude this section, we want to discuss the assumptions underlying a very well know model which has been widely used for estimating the value of the edge tension $\gamma$ from experiments \cite{Brochard(2000)}. This model is based on the following equation for the pore radius
\begin{equation}
\frac{dr}{dt}= 2\pi L_{r r}\left\{\sigma r-\gamma\right\}
\label{eq:ec_dotrpApp}
\end{equation}
and on the following equation for the vesicle radius
\begin{equation}
\dot{R}=-\frac{2\sigma}{3\eta_{s}} \dfrac{r^{3}}{R^{3}}.
\label{eq:ec_DPLaplace}
\end{equation}
In writing this system of equations one has to assume that the vesicle free energy does not depends on the vesicle radius $R$ and that the total surface energy only depends on the surface tension $\sigma$ [given by Eq. (\ref{eq:ec_SigAp})], that is, the bending contribution $\kappa_b$ is neglected. In addition, the equilibrium Laplace relation between the surface tension and the pressure difference is assumed to hold during the whole process, $\Delta P = 2\sigma/R$. As a consequence of these assumptions, the model implicitly assumes that $L_{rR}=0$ (no cross effects) and that $L^{-1}_{rr}= 4 \pi \eta_2 d$, with $\eta_2$ the membrane viscosity (which is an unknown parameter) and $d$ the bilayer width \cite{Brochard(2000)}.
 \begin{table}
\caption{Physical quantities used in the fittings of the experimental data reported in \cite{ErdemK(2003),ThomasP(2010),Brochard(2000)}.} 
\label{tab:Table1}       
\begin{center}\tabcolsep=7.0pt
\begin{tabular}{@{}lcccccr@{}}
\hline\noalign{\smallskip}
$\eta_{s}$(cP)& $R_{eq}$($\mu$m) & $r_{0}$($\mu$m)&$R_{0}$($\mu$m)&$\kappa_{b}$($10^{-20}$J) & $\sigma_{c}$($10^{-5}$N$\cdot$m) & $\gamma$(pN)\\
\noalign{\smallskip}\hline\noalign{\smallskip}\hline\noalign{\smallskip}
32 &19.6 &  1.10 & 20.5 &29 & 2.9 & 0.92 \\
1.13 &19.7 & 1.15 & 20.4 & 29 & 3.0 & 1.21 \\
32 &14.5 & 1.15 & 14.7 & 29 & 3.1 & 0.57 \\ 
\noalign{\smallskip}\hline
\end{tabular}
\end{center}
\end{table}
Eqs. (\ref{eq:ec_dotrpApp}) and (\ref{eq:ec_DPLaplace}) reproduce the experimental results provided that the vesicle radius dynamics, $dR/dt$, is controlled by the solution viscosity $\eta_{s}$ whereas the membrane viscosity, 
$\eta_{2}$, dictates the pore dynamics. This fact seems to be inconsistent with experimental evidence \cite{ErdemK(2003),ThomasP(2010)}. In addition, Eq. (\ref{eq:ec_dotrpApp}) does not satisfies the condition that it must be a critical point, that is, $dr/dt=0$ and $dR/dt=0$ at $r=0$ and $R=0$. From this equation it is clear that $dr/dt|_{r=0}\neq 0 $ yielding to negative the pore radius.

\section{Discussion and conclusions}\label{sec:concl}
In this work, we propose a theoretical model that accounts for all the features of the dynamics of transient pores in spherical lipid vesicles, see Figures \ref{fig:Dyna_Karatekin}(a),  \ref{fig:Dyna_Dimova}(a) and \ref{fig:Dyna_DeGennes}(a). Essentially, it allows to determine the value of pore edge tension along with the time course of the vesicle radius, the effective surface tension and the pressure difference from the measurements of the pore radius \cite{ErdemK(2003),ThomasP(2010)}, see Figures \ref{fig:Dyna_Karatekin}(b)-(d),  \ref{fig:Dyna_Dimova}(b)-(d) and \ref{fig:Dyna_DeGennes}(b)-(d). All the parameters included have a clear physical meaning and can be directly measured in experiments; for example, the viscosity of the solution, the bending modulus or the initial surface tension, see Table \ref{tab:Table1}.

The model was deduced by using Onsager's irreversible thermodynamics and it consists on two coupled first-order time differential equations for the pore and the vesicle radii, $r(t)$ and $R(t)$ respectively. The coupling emerges from the dependence of the surface tension and the pressure difference on the state variables ($r$ and $R$) and from cross effects coming from the linear laws of non-equilibrium thermodynamics. The latter introduce Onsager's phenomenological coefficients whose dependence on $r$ and $R$ is provided to have a consistent physical model. This condition is not fulfilled by previous models  \cite{Brochard(2000),RRyham(2011)}. In agreement with experiments, 
the characteristic relaxation time of the pore dynamics is controlled by the solution viscosity, $\tau_{relax} \propto \, \eta_{s}$, see Eqs. (\ref{eq:ec_rpAll}) and (\ref{eq:ec_RVAll}). 

The continuous lines in \ref{fig:Dyna_Karatekin}(b),  \ref{fig:Dyna_Dimova}(b) and \ref{fig:Dyna_DeGennes}(b) show the evolution of the radius of the vesicle, as predicted by the numerical solution of Eqs. (\ref{eq:ec_rpAll}) and (\ref{eq:ec_RVAll}). This behavior is in agreement with the experimental results, Fig. \ref{fig:Dyna_Karatekin}(b). In particular, in experiments with constant illumination, this evolution is typical, and the vesicle radius may be reduced by 40\% \cite{NRodriguez(2006)}. 
The panels (c) and (d) of Figures \ref{fig:Dyna_Karatekin}, \ref{fig:Dyna_Dimova} and \ref{fig:Dyna_DeGennes} show that after the pore formation, the surface tension and the pressure difference decrease drastically until the pore radius reaches its maximum value. After that, both increase slowly until a final equilibrium value of about $25\%$ of their initial value. 

The previous results show that the membrane is not fully relaxed after the pore collapse. Furthermore, correlating the results shown in Figures \ref{fig:Dyna_Karatekin}(a), \ref{fig:Dyna_Dimova}(a) and \ref{fig:Dyna_DeGennes}(a) with their counterparts in (c) and (d), and with Eqs. (\ref{eq:ec_rpAll}) and (\ref{eq:ec_RVAll}), we see that it is more convenient to calculate the edge tension, 
$\gamma$, from the slope of the  initial quasi-linear relaxation regime, because the change of the slope in the final collapse stage manifests clearly the non-linear nature of the process attributable to the interplay among the edge and the surface tensions together with the pressure difference. In Fig. \ref{fig:Dyna_Karatekin}(a) we show this through the average slopes of the two regimes, shown by the solid and dashed straight lines. 

Near the initial quasi-linear relaxation regime, both $\tilde{\sigma}$ and $\Delta P$ nearly vanish [see Figs. 2(a)- 4(a)] and, therefore, Eqs. (\ref{eq:ec_rpAll}) and (\ref{eq:ec_RVAll}) may be approximated to give the following relaxation equation for the pore radius
\begin{eqnarray}
\frac{dr}{dt}=-\dfrac{\gamma}{r_{0}\eta_{s}R_0} r.
\label{eq:ec_rfinal}
\end{eqnarray}
From this equation we can obtain a simple and useful expression for the edge tension as a function of the pore radius 
\begin{eqnarray}
\gamma \simeq - r_{0}\eta_{s}R_0 \frac{\Delta\ln{|r|}}{\Delta t}.
\label{eq:ec_rfinal-diferenecias}
\end{eqnarray}
Using this expression we can calculate the values of the edge tensions reported in the first column of Table 2.  These results correlate well with those predicted by the whole model that incorporates all the mechanisms participating in the relaxation dynamics; that is, it includes the effects due to the surface tension and the pressure difference. 
If Eq. (\ref{eq:ec_rfinal-diferenecias}) is used in the final regime (a rough approximation) then the values obtained for the edge tension are about one order of magnitude larger than in the first regime. Notice, in addition, that these values almost duplicate the corresponding ones reported in previous works \cite{ErdemK(2003),ThomasP(2010),Brochard(2000)}, see the third and fourth columns of Table 2. The reason is that previous estimations of the edge tension have been obtained using the model proposed in Ref. \cite{Brochard(2000)}. As discussed after Eqs. (\ref{eq:ec_dotrpApp}) and (\ref{eq:ec_DPLaplace}), the election of the Onsager coefficients in that case corresponds to $L^{-1}_{rr} \sim 4 \pi \eta_2 d$ whereas in our case should be $L_{rr}^{-1} \sim 2\pi r_{0} \eta_{s}$, involving a factor of two. 
\begin{table}[ht!]
\caption{Comparison of the edge tensions $\gamma$ obtained in the different relaxation regimes and with previous results reported.} 
\label{tab:Table2}       
\begin{center}\tabcolsep=7.0pt
\begin{tabular}{@{}lcccr@{}}
\hline\noalign{\smallskip}
Figure & Initial regime & Whole model & Final  regime & Previous results\\
\noalign{\smallskip}\hline\noalign{\smallskip}\hline\noalign{\smallskip}
2 & 1.60 pN  & 0.92 pN &  26.5 pN  & 10.0 pN, Ref. \cite{ErdemK(2003)}\\
3 & 1.32 pN & 1.21 pN & 14.3 pN  & 14.3 pN, Ref. \cite{ThomasP(2010)}\\
4 & 1.01 pN & 0.57 pN & 10.8 pN   & 6.0 pN, Ref. \cite{Brochard(2000)}\\ 
\noalign{\smallskip}\hline
\end{tabular}
\end{center}
\end{table}

The values of the parameters taken from experimental data reported in the literature are the solvent viscosity, the initial and final vesicle radius, the characteristic value of the surface tension and the bending modulus \cite{ErdemK(2003),ThomasP(2010),Brochard(2000),WRawicz(2000),Lipowsky(1995)}. The magnitude of the initial pore radius can be obtained from the relation for the surface tension. Therefore, all fits were obtained by only adjusting the edge tension.

It is convenient to mention that pore formation in vesicles immersed in high viscosity solutions (32cP), Figs. 2 and 4, is about $\delta t_{formation}= 0.1-0.2$ seconds. In contrast, when the solution's viscosity is near that of the cytoplasm (bewteen 1-3 cP, as in Figure 3), then the formation time is much less, 
$\delta t_{formation}= 0.0075$ seconds. This second formation time scale is in the order of magnitude with what is expected for pore dynamics in living cells. It is interesting to notice that for the low viscosity experiment, Figure 3, all the values reported in Table 2 are consistent between them, in contrast with the data obtained in the other two cases.

To conclude, we may stress that Onsager's irreversible thermodynamics is a powerful formalism that allows to deduce simple and consistent physicochemical models for biological small systems, in particular, to describe the dynamics of pores in biological membranes. The obtained equations are consistent with the second law of the second law of Thermodynamics and it may easily be generalized to more complex situations like those in which noise or external forces affects the behavior of the system.

\begin{acknowledgements}
We thank  to A. Ledesma-Dur\'an and A. Arteaga-Jim\'enez for their useful discussions. This work was supported by UNAM DGAPA Grant No. IN113415 and CONACYT. 
\end{acknowledgements}

\end{document}